\newcommand{\ndy}{(Nd$_x$Y$_{1-x}$)$_{2}$BaNiO$_{5}$}
\newcommand{\nd}{Nd$_{2}$BaNiO$_{5}$}
\newcommand{\pr}{Pr$_{2}$BaNiO$_{5}$}
\newcommand{\er}{Er$_{2}$BaNiO$_{5}$}
\newcommand{\ho}{Ho$_{2}$BaNiO$_{5}$}
\newcommand{\y}{Y$_{2}$BaNiO$_{5}$}
\newcommand{\rr}{$R_{2}$BaNiO$_{5}$}

\documentstyle[aps,rmp,twocolumn]{revtex}
\begin{document}
\title{Quantum and classical dynamics in complex one-dimensional
antiferromagnets.}
\author{A. Zheludev and S. Maslov}
\address{Physics Department, Brookhaven National Laboratory, Upton, NY 11973,
USA.}
\author{ T. Yokoo and J. Akimitsu}
\address{Department of Physics, Aoyama-Gakuin University, 6-16-1,
Chitosedai, Setagaya-ku, Tokyo 157, Japan.}
\author{S. Raymond}
\address{ DRFMC/SPSMS/MDN, CENG, 17 rue des Martyrs, 38054
Grenoble Cedex, France.}
\author{S. E. Nagler}
\address{Oak Ridge
National Laboratory, Bld. 7692, MS 6393, P.O. Box 2008, Oak Ridge, TN 37831,
USA.}
\maketitle

\begin{abstract}
Of great recent interest in condensed matter physics are phenomena
of coexistence of quantum and classical properties in the same
material. Such duality occurs in certain mixed-spin
antiferromagnets composed of quantum spin chains interacting
through ``auxiliary'' magnetic ions. In this category are
linear-chain rare earth nickelates that exhibit a seemingly
paradoxical co-existence of long-range magnetic order and
one-dimensional gapped quantum spin excitations at low
temperatures. In the present paper we propose a unified physical
description of these compounds, supported by recent neutron
diffraction and inelastic scattering studies. Our basic concept is
the effective separation between low-frequency classical and
high-frequency quantum spin correlations. This interpretation
enables experimental measurements of some previously inaccessible
fundamental properties of quantum spin chains, and predicts new
exotic magnetic excitations and  mechanisms of long-range ordering
in complex quantum magnets.
\end{abstract}

\tableofcontents

\section{INTRODUCTION}
\label{sec:introduc} For the past two decades one-dimensional (1D)
quantum antiferromagnets (AFs) have been among the hot topics in
condensed matter physics. Conventional three-dimensional magnetic
materials exhibit long-range magnetic order (LRO) at low
temperatures. In contrast, in one- and two-dimensional systems
thermal fluctuations destroy LRO at any temperature $T>0$. The 1D
AF case is particularly interesting since even at $T=0$ LRO is
destroyed by {\it zero-point quantum fluctuations}. A breakthrough
in understanding quantum AFs came with the theoretical work of
Haldane (Haldane, 1983), who showed that spin correlations are
radically different in integer- and half-integer 1D spin systems.
Half-integer-spin chains are known to be in a {\it quantum
critical state}, where spin correlations decay with distance
according to a power law, and the excitation spectrum is gapless.
In contrast, the ground state of integer spin chains is a
``quantum spin liquid'', with exponentially decaying correlations
and an energy gap $\Delta$ in the magnetic excitation spectrum.

The wealth of experimental and theoretical results accumulated to
date adds up to a fairly complete understanding of many of the
properties of idealized 1D quantum AFs. One remaining problem of
enormous interest is understanding {\it quasi}-1D systems:
interacting quantum spin chains as part of {\it real}
3-dimensional crystal structures. Significant insight in this
field has been obtained in studies of simple compounds with {\it
directly coupled} quantum spin chains, such as CsNiCl$_3$ ($S=1$)
(see, for example, Buyers {\it et al.,}, 1986, Morra {\it et al.},
1988, Enderle {\it et al.}, 1999, and references therein) and
KCuF$_3$ ($S=1/2$) (Tennant {\it et al.}, 1993, Tennant {\it et
al.}, 1995, Lake {\it et al}, 2000, and references therein).
Quasi-1D AFs are a particular case of the more general and very
interesting phenomenon of the coexistence of quantum and classical
spin fluctuations in the same material. In this context there have
been a number of studies of classical systems with an added
component of quantum mechanical interactions. An example of this
is the realization of the so-called ``Transverse Ising Model''
recently found in certain holmium salts (Brooke {\it et al.},
1999). In the present paper we discuss the opposite approach,
starting from the quantum limit, and looking at quantum spin
chains that interact with each other only indirectly, via an
``auxiliary'' network of essentially classical spins.  The only
known experimental realizations of this model are rare earth
nickelates with the general formula \rr, where $R$ stands for a
magnetic rare earth element. These ``mixed'' quantum-classical
systems contain almost isotropic $S=1$ AF Ni-chains weakly coupled
to magnetic rare earth ions. As discussed below, they exhibit an
effective {\it separation} of classical and quantum spin dynamics
in certain regimes. This provides a unique opportunity to
investigate a number of interesting phenomena, including the
effect of strong staggered magnetic fields on Haldane spin chains,
and the peculiar interaction between collective quantum
excitations in the chains and single-ion quantum transitions in
the rare earth elements.

All \rr\ species are derivatives of \y, an isomorphous compound in
which the magnetic $R^{3+}$ are replaced by non-magnetic Y$^{3+}$
ions. \y\ is a textbook example of a Haldane-gap  AF (Darriet and
Regnault, 1993, Yokoo {\it et al.,} 1995, Xu {\it et al.}, 1996).
Its magnetic properties are due to chains of $S=1$ Ni$^{2+}$ ions
that run along the $a$ axis of the orthorhombic crystal structure
(Fig.~\ref{struct}). The Ni-spins are coupled by rather strong
($J\sim 300$~K) AF nearest-neighbor intra-chain interactions.
Inter-chain coupling is negligible. Down to very low temperatures
this system shows no sign of magnetic ordering and has a
thermally-activated magnetic susceptibility: a signature of a
spin-singlet ground state and energy gap. Haldane-gap excitations
in \y\ ($\Delta\approx 10$~meV)  have been observed and
extensively studied in inelastic neutron scattering experiments
(Darriet and Regnault, 1993, Yokoo {\it et al.}, 1995, Xu {\it et
al.}, 1996). In \rr\ compounds, the place of Y$^{3+}$ is taken by
magnetic rare earth  ions ranging from Pr$^{3+}$ to Tm$^{3+}$
(GarciaGarc\'{i}a-Matres {\it et al.}, 1993,
GarciaGarc\'{i}a-Matres {\it et al.}, 1995). Even though direct
interactions between the $R^{3+}$ moments are negligible, their
coupling to the Ni-chains in all cases leads to long-range
N\'{e}el ordering with transition temperatures $T_{\rm N}$ ranging
from 24~K (\pr) to 65~K (Dy$_2$BaNiO$_5$). A static ordered moment
appears on both the Ni- and $R$-sites simultaneously upon cooling
through $T_{\rm N}$ (Figs.~\ref{MvsH}a,b). At a first glance it
may seem that the magnetic LRO is a clear sign of a destruction of
the singlet ground state in individual Haldane chains. Indeed, for
quite some time \rr\ compounds were considered to be classical
magnets.

A breakthrough came when, in inelastic neutron scattering
experiments on \pr\ (Zheludev {\it et al.}, 1996-1), \nd (Zheludev
{\it et al.}, 1996-2, Yokoo {\it et al.}, 1998, Raymond {\it et
al.}, 1999, Zheludev {\it et al.}, 2000), and \ndy\ (Yokoo {\it et
al.}, 1997, Yokoo {\it et al.} , 1998), it was discovered that 1D
Ni-chain gap excitations, strikingly similar to the Haldane modes
in \y, not only exist above the ordering temperature, but also
{\it persist in the magnetically ordered state}. One could argue
that these excitations in \rr\ are classical spin waves, that
acquire a gap due to magnetic anisotropy effects, and that the
similarity with \y\ is coincidental. This interpretation, however,
is clearly inappropriate. The gap energy (around 10~meV) is too
large to be attributed to anisotropy on the Ni-sites (less than
1~meV, according to Xu {\it al.}, 1996). At the same time, it can
not be attributed to anisotropy involving the rare earth ions
either, being independent of the type of rare earth involved, and
even remaining unchanged in the entire concentration range of
mixed Y-$R$ compounds. Another hint is that classical spin waves
represent the spontaneous breaking of $O(3)$ symmetry in the
magnetically ordered state, and are thus a doublet of excitations.
In contrast, gap excitations in \rr\ were shown to be a {\it
triplet} (Raymond {\it et al.}, 1999), just like Haldane
excitations from an $O(3)$-invariant spin liquid state. One is
forced to conclude that the 1D quantum spin correlations in the
Haldane chains somehow survive the onset of LRO in \rr\ compounds.

\section{TWO FREQUENCY REGIMES}
\label{sec:regimes}
The key to resolving this apparent contradiction lies in the very short time
scale associated with dynamic spin correlations in individual chains, which in
turn is a {\it direct consequence of the energy gap}. At frequencies
substantially smaller than the threshold frequency $\Delta/\hbar$, the dynamic
susceptibility $\chi(\omega)$ of an isolated $S=1$ AF chain is purely real and
is {\it only weakly frequency-dependent}. In \rr\ the magnitude $J_{\bot}$ of
Ni-$R$ exchange interactions is typical of a rare-earth-transition-metal
superexchange bond, and is of the order of 1~meV.  This energy and
corresponding frequency scales are much smaller than the Haldane gap $\Delta$
(10~meV). At such frequencies the local magnetization of the Ni-chains is able
to instantaneously follow any fluctuations of neighboring rare earth spins. The
effect of the quantum spin chains is then reduced to simply providing an
effective pathway for $R$-$R$ interactions of the magnitude $J_{\rm eff}\sim
J_{\bot}^2 \chi(0)$. We end up with a 3D network of essentially classical rare
earth moments that, not surprisingly, orders magnetically at low temperatures.
The low-frequency dynamics are then also that of a classical magnet.
Conversely, the same arguments suggest that the dynamic spin correlations in
the high-frequency range are those of a 1-D $S=1$ quantum AF in a {\it static}
effective field generated by the rare earth subsystem. In other words, in the
limit of weak coupling, for two-component systems that include {\it gapped}
quantum spin chains there is a {\it separation} between static and
low-frequency properties (LRO and classical spin waves) and the high-frequency
behavior (quantum 1-D excitations). In the next two sections, we shall review
the experimental and theoretical results that support this physical picture.

\section{STATIC PROPERTIES}
\label{sec:static} A quantitative model that embodies the
hand-waving arguments given above is the chain-mean-field
(chain-MF) theory (Scalapino {\it et al}, 1975, Affleck, 1989).
The standard MF approach for conventional magnets starts out with
the bare (non-interacting) susceptibilities for all individual
magnetic ions involved. In the chain-MF theory, one respects the
fact that Ni-$R$ coupling is substantially smaller than the strong
interactions within the Ni-chains themselves. The chains are thus
treated as single entities. For the \rr\ series the
low-temperature magnetic structure is antiferromagnetic, so the
relevant properties of the spin chains are their {\it staggered
magnetization function} $M_{\pi}(H_{\pi})$ and {\it staggered
susceptibility} $\chi_{\pi}=dM_{\pi}/dH_{\pi}$. As for the rare
earth ions, for those with a Kramers doublet ground state
configuration (Nd$^{3+}$ and Er$^{3+}$ among them) it is
appropriate to take the standard approach and write the
non-interacting response in the form of a Brillouin function for
an isolated ion. The coupling between these two systems is then
treated at the MF level and all static properties, including
long-range ordering and the $T$-dependence of the magnetic order
parameters are derived from a set of self-consistent equations.
The problem is that to solve these equations one has to know
$M_{\pi}(H_{\pi})$ for an isolated chain. Prior to the
experimental studies of \rr\ this function was not known
analytically or even numerically. Simple powder-diffraction
measurements on \ndy, \er\ and \ho\ however, allowed the direct
{\it measurement} of this fundamental property (Zheludev {\it et
al.}, 1998, Yokoo {\it et al.}, 1998). The magnetic ordering
temperatures (around 50~K) are considerably smaller than the
Haldane gap energy ($\Delta/k_{\rm B}\sim120$~K) and, in this
regime, $M_{\pi}(H_{\pi})$ is almost $T$-independent (Kim, 1998).
The effective staggered field acting on the spin chains in \rr\ is
proportional to the rare earth magnetic order parameter. From
these two facts it follows  that $M_{\pi}(H_{\pi})$ may be
obtained simply by plotting the measured ordered moment on the
Ni-sites vs. that on the $R$-sites, as shown in
Fig.~\ref{MvsH}(c). The scaling of the abscissa depends on the MF
coupling constant and obviously on the type of rare earth ions
involved. Universal scaling in proper magnetic field units is
achieved by substituting the measured $M_{\pi}(H_{\pi})$ curve
into the self-consistent MF equations. The MF coupling constants
are then refined by fitting the measured temperature dependencies
of sublattice magnetizations (Zheludev {\it et al.}, 1998, Yokoo
{\it et al.}, 1998). Excellent fits can be obtained in this manner
(Fig.~\ref{MvsH}a,b- solid lines).  The resulting field scale is
shown in the top axis of Fig.~\ref{MvsH}(c). The data collapse for
systems with substantially different ordering temperatures and
saturation moments is quite impressive, validating this approach.
Note that staggered fields of up to 40~T are being produced in
these experiments!

It is important to realize that the mechanism of magnetic ordering
in \rr\ materials is quite different from that in systems with
directly coupled chains, such as CsNiCl$_3$. In the latter, LRO
results from a complete softening of the Haldane excitations at
the 3-D AF zone-center as $T\rightarrow T_{\rm N}$. This can
happen only if inter-chain interactions exceed some critical value
(Affleck, 1989). In the case of rare earth nickelates, magnetic
ordering is expected to occur for any {\it arbitrarily small}
Ni-$R$ coupling, as a result of the $1/T$-divergence in the bare
susceptibility of individual rare earth moments.

\section{SPIN DYNAMICS}
\label{sec:dynamic}

While long-range magnetic ordering is beautifully described by the
chain-MF model, the spin {\it dynamics} in our two-component
magnets can be analyzed with the chain-Random-Phase-Approximation
(RPA). At this level, just as was the case for the static
properties, the low-energy magnetic excitations in \rr\ can be
described in terms of effectively coupled classical rare earth
spins. Such classical spin waves were recently studied in \nd\
(Yokoo {\it et al.}, 1998). The Nd-case is particularly simple
because of a strong easy-axis magnetic anisotropy associated with
Nd$^{3+}$. The spin waves are then dispersionless and resemble
excitations in a conventional Ising magnet. The excitation energy
is given by $2M_{\rm Nd}H_\pi$, where $M_{\rm Nd}$ is the
saturation moment of Nd$^{3+}$ and $H_\pi$ is the effective mean
field produced by the Ni-sublattice. In fact, the measured
temperature dependence of this energy can be accurately
reproduced, without using any adjustable parameters, by utilizing
$H_\pi(T)$ determined from the MF-analysis of the sublattice
magnetization data.

We now turn to the Ni-chain gap excitations. In the paramagnetic
phase (above $T_{\rm N}$) there is no ordered moment on the rare
earth sublattice, so the 1-D gap modes are exactly as those in
uncoupled chains. Thus in \nd\ and \pr, for $T>T_{\rm N}$, the 1-D
gap excitations are experimentally identical to those found in \y\
(Zheludev {\it et al.}, 1996-1), \nd (Zheludev {\it et al.},
1996-2, Yokoo {\it et al.}, 1998, Raymond {\it et al.}, 1999). In
particular, the $T$-dependence of the gap energy is very similar
(Fig.~\ref{gap}a). Within the framework of our MF-RPA model, in
the magnetically {\it ordered} state, the gap excitations are
those of isolated quantum spin chains immersed in a {\it static}
staggered exchange field. Experimentally, Haldane-gap modes are
indeed observed for $T<T_{\rm N}$ in all \rr\ materials studied to
date. The striking experimental result is that the gap energy {\it
increases} with decreasing $T$ below $T_{\rm N}$, roughly linearly
with $T-T_{N}$ (Fig.~\ref{gap}a). Moreover, the behavior appears
to be independent of the type of rare earth involved or the actual
N\'{e}el temperature. The best way to demonstrate this
universality is to eliminate the rare-earth related energy scale
completely by plotting the increase of the Haldane gap, normalized
to that in \y, as a function of the induced static staggered
moment on the Ni-sites (Fig.~\ref{gap}b). When plotted in this
way, the data obtained for different \rr\ systems collapses onto a
single curve. This is a clear indication that an increase of the
Haldane gap energy in the presence of a staggered field is an {\it
intrinsic} property of the quantum spin chains.

Despite the simplicity of its origin, the effective separation of
classical and quantum dynamics in these materials has a powerful
consequence: it turns the \rr\ compounds into unique model systems
for fundamental studies of quantum spin chains in strong staggered
fields, {\it i.e.}, fields modulated on the {\it atomic} length
scale. The direct measurements of the staggered magnetization
curve $M_{\pi}(H_{\pi})$ and the staggered field dependence of the
Haldane gap $\Delta(H_\pi)$ described above, stimulated intensive
theoretical studies. A particularly successful approach (Maslov
and Zheludev, 1998, Yokoo {\it et al.}, 1998) is the
$O(3)$-symmetric (1+1)-dimensional field theory (the so-called
$\phi^4$ model), first used by Affleck to describe Haldane spin
chains (Affleck, 1989). In this theory it becomes apparent that
the non-linearity of $M_{\pi}(H_{\pi})$ and the monotonous
behavior of $\Delta(M_\pi)$ are both a manifestation of {\it
repulsion} between single-particle gap excitations. Coefficients
characterizing this repulsion in a fully renormalized Hamiltonian
can be estimated numerically and used to obtain power series
expansions for $M_{\pi}(H_{\pi})$ and $\Delta(M_\pi)$ (Maslov and
Zheludev, 1998). These predictions are in excellent agreement with
all of the \rr\ data and are shown as solid lines in
Figs.~\ref{gap}b and Fig.~\ref{MvsH}c. The results obtained in
neutron scattering experiments on \rr\ have also been confirmed by
numerical simulations (Lou {\it et al.}, 1999, dashed line in
Fig.~\ref{gap}b). Very recently, the increase of the gap energy in
a two-component magnet was rigorously proven for the Valency Bond
Solid model (Bose and Chattopadhyay, 1999), known to possess many
similarities with 1-D $S=1$ Heisenberg model relevant to the
present case.

\section{THE ROLE OF SINGLE-ION EXCITATIONS}
\label{sec:singleion}

Above we restricted ourselves to considering Haldane chains coupled to
essentially classical spins. Qualitatively new phenomena occur when the
``auxiliary'' magnetic ions themselves have non-trivial intrinsic dynamics. Due
to strong spin-orbit interactions and low site-symmetry, this in fact is always
the case for $R^{3+}$ in \rr. Not just the ground state multiplets of the rare
earth ions, but also their higher-energy excited crystal-field (CF)
configurations will couple to the integer-spin chains.

One can identify two distinct regimes. The first regime is
realized when the $R$-centered single-ion CF transitions occur
close to, or above, the Haldane gap. CF excitations in this case
have little influence on the low-frequency and static properties
of the system, for which the classical model remains adequate. At
high frequencies though, the  proximity of single-ion modes on the
$R$ sites to 1-D excitations in the chains gives rise to peculiar
mixed Ni-$R$ modes. Such behavior was recently observed in
inelastic neutron scattering experiments on \nd (Zheludev {\it et
al.}, 2000). Figures~\ref{exdata}(a,b) show representative
constant-$Q$ scans collected at the 1-D AF zone-center for
different momentum transfers perpendicular to the chain axis, at
$T=55$~K ({\it above} $T_{\rm N}=48$~K). Three separate peaks,
namely the Haldane-gap excitation at 11~meV energy transfer and
two CF excitations, a weaker mode  at 18~meV and  a strong one at
24~meV, can be clearly identified. A striking feature is the
strong variation of the intensity of these modes as a function of
transverse momentum transfer, as shown in (Fig.~\ref{exdata}c).
This intensity modulation results from an {\it interference}
between Ni- and Nd- spin fluctuations. The gap excitations retain
their predominantly 1-D character, and can be described as Haldane
modes propagating on the Ni-chains, ``dressed'' by correlated CF
fluctuations on the Nd-sites. This behavior is well accounted for
by a slightly more sophisticated version of the chain-RPA model,
that includes higher-energy levels of the rare earth ions. It is
interesting that at the 1-D AF zone-center the interference effect
barely influences the energies of the modes coupled. Only a very
weak transverse dispersion is seen in the Haldane branch
(Fig.~\ref{exdata}d), and none can be detected in the CF
excitations within experimental energy resolution. Our
interpretation of the $T$-dependence of the gap in terms of
Haldane chains in a static staggered field is thus justified. It
is also clear that the Ni-Nd interactions in \nd\ are too weak to
drive the Haldane gap to zero energy at any wave vector, which
would be required to produce a CsNiCl$_3$-like transition to a
N\'{e}el-like state (Buyers {\it et al.}, 1986, Morra {\it et
al.}, 1988, Affleck, 1989).

Away from the 1-D AF zone-center the mixing between Haldane excitations and
local modes can be quite dramatic in the isolated regions of reciprocal space
where their non-interacting energies coincide. Here both intensities and
energies of coupled modes are severely affected. In fact, the distinction
between single-ion and chain modes is totally blurred: the former continuously
``morph'' into the latter, and vise versa. This ``anticrossing'' phenomenon in
\nd\ is illustrated in Fig.~\ref{exdata2}, that shows a mesh composed of data
points collected in constant-$E$ scans along the chain axis.

Quite different is the second regime, where the energies of
certain excited CF states are considerably smaller than the
Haldane gap. As in the case of classical ``auxhiliary'' spins,
rare earth- and Ni-chain- spin dynamics are effectively separated.
At low frequencies the Haldane chains simply provide effective
pathways for direct $R$-$R$ interactions.  This effective coupling
becomes particularly important for non-Kramers rare earth ions
that in the structure of \rr\ have a non-magnetic ground state. A
good example is \pr, where Pr$^{3+}$ has a spin-singlet ground
state, but in addition has a CF-excited state at only 4~meV
energy. Above $T=100$~K the corresponding CF mode is
dispersionless. With decreasing temperature it acquires a
substantial dispersion and intensity variation both along, and
perpendicular to, the chain axis (Zheludev {\it et al.}, 1996-1).
With two Pr sites per every Ni-site, The CF mode becomes split
into an optic and accoustic branches. As shown in Fig.~\ref{Pr},
the accoustic branch develops a sharp dip at the 3D AF
zone-center, where its energy reaches zero at $T_{\rm N}=24$~K. A
magnetic {\it soft-mode} transition occurs at this point, and is
driven by Ni-chain-mediated Pr-Pr interactions. This soft-mode
behavior is very similar to that in metallic Pr (for a good review
of the subject, see book by Jensen and Mackintosh). Below $T_{\rm
N}$ \pr\ is ordered in a structure similar to that of \nd. In an
unusual twist, magnetic LRO in this case is a result of
interaction between two {\it non-magnetic} systems:
singlet-ground-state rare-earth ions and quantum-disordered spin
chains.

\noindent
{\bf $\dots$}

In summary, complex systems composed of gapped quantum spin chains interacting
through isolated free spins demonstrate a rich spectrum of magnetic properties
with both classical and quantum-mechanical features. $R_2$BaNiO$_5$ rare earth
nickelates are unique model compounds in which these phenomena can be studied
experimentally.

\section*{ACKNOWLEDGMENTS}

Work at BNL was carried out under Contract No. DE-AC02-98CH10886, Division of
Material Science, U.S. Department of Energy. Oak Ridge National Laboratory is
managed for the U.S. D.O.E. by Lockheed Martin Energy Research Corporation
under contract DE-AC05-96OR22464.

\pagebreak

\pagebreak
\begin{figure}
\caption{A schematic view of the \rr\ crystal structure. $S=1$
Haldane spin chains are formed by Ni$^{2+}$ ions at the centers of
NiO$_6$ octahedra , arranged along the $a$ crystallographic axis.
The rare earth ions provide magnetic links between chains.}
\label{struct}
\end{figure}
\begin{figure}
\caption{(a) Temperature dependence of the Ni- and R-sublattice
magnetizations measured in \nd\ (symbols) (Yokoo {\it et al.},
1998) using neutron powder diffraction. The solid lines are
chain-MF theoretical fits. (b) Same for \er (Alonso {\it et al.},
1990). (c) Measured induced staggered moment on the Ni-sites
(symbols) plotted against the magnetic order parameter of the
$R$-sublattice in several \rr\ compounds (bottom axes). The
collapsed data sets can be interpreted as a measurement of the
universal staggered magnetization curve for Haldane spin chains
(top axis). The solid line is a theoretical prediction based on
the $\phi^4$ model.} \label{MvsH}
\end{figure}
\begin{figure}
\caption{ (a) Measured temperature dependence of the Haldane gap
energy for a series of Y-substituted \nd\ compounds (symbols), as
measured by inelastic neutron scattering. Above the corresponding
temperatures of magnetic ordering all systems show the same
behavior as the intrinsically disordered \y\ (Yokoo {\it et al.},
995) . In the ordered phase the gap increases in all cases
linearly with $T$. Lines are guides for the eye. (b) Data collapse
obtained by plotting the measured difference between  the gap in
\ndy\ and that in \y\ as a function of the induced staggered
moment on the Ni-chains. The solid line is a parameter-free
prediction of the $\phi^4$ model, and the dashed line represents
recent numerical results (Lou {\it et al.}, 1999) for Haldane
chains in a staggered field.} \label{gap}
\end{figure}
\begin{figure} \caption{Mixing between Haldane-gap and crystal field
excitations in \nd, observed in single-crystal inelastic neutron
scattering experiments. (a,b) Typical constant-$Q$ scans measured
at the 1D AF zone-center for two different momentum transfers
perpendicular the chain-axis (symbols). The data were taken on the
HB-1 3-axis spectrometer at Oak Ridge National Laboratory, using
$E_f=14.7$~meV fixed-final energy neutrons, pyrolitic graphite
(PG) monochromator and analyzer,  $60'-80'-80'-240'$ collimations
and a PG filter after the sample. The solid black line is a fit to
a semi-empirical model cross section taking into account
resolution effects. The red, green and blue peaks represent the
Haldane and two CF excitations, respectively. This analysis
reveals a substantial intensity modulation in all modes (c) and a
weak dispersion in the Haldane mode (d).} \label{exdata}
\end{figure}
\begin{figure}
\caption{Grayscale/contour plot of inelastic neutron scattering
intensity measured in \nd\ at the 1D AF zone-center  shows the
highly dispersive Haldane-gap mode (dashed parabola) and the flat
24~meV crystal field mode (dashed horizontal line). The data were
taken using the same setup as for Fig.~4. The symbols and attached
bars show the positions and widths of peaks seen in constant-$E$
(circles) and constant-$Q$ (squares) scans, respectively. A mixing
between the two modes leads to an anticrossing effect (solid
lines).} \label{exdata2}
\end{figure}
\begin{figure}
\caption{Temperature dependence of the 3D AF zone-center
excitation energies measured in Pr$_2$BaNiO$_5$. A repulsion
between single-ion Pr CF excitations (circles) and Haldane-gap
modes in the Ni-chains (triangles) leads to a soft-mode transition
to a  magnetically ordered state below $T_{\rm N}=24$~K, despite
the singlet (non-magnetic) nature of the Pr$^{3+}$ ions. The
Haldane-gap modes behave very similarly to those in \nd. }
\label{Pr}
\end{figure}


\begin{thebibliography}{10}

\bibitem{Affleck89-2}
Affleck, I., Model for quasi-one-dimensional antiferromagnets:
Application to CsNiCl$_3$. {\it  Phys. Rev. Lett.} {\bf 62},
474-477  (1989).

\bibitem{Alonso90}
Alonso, J. A., J. Amador, J. L. Mart\'{i}nez, I. Rasines, J.
Rodr\'{i}guez-Carjaval and R. S\'{a}ez-Puche, Neutron diffraction
study of the magnetic structure of Er$_2$BaNiO$_5$. {\it  Solid
State Comm.} {\bf 76}, 467--474 (1990).

\bibitem{Bose99}
Bose, I.  and E. Chattopadhyay, Mixed-spin systems: coexistence of
Haldane gap and antiferromagnetic long range order, Phys. Rev. B
{\bf 60},  12138 (1999).


\bibitem{Brooke99}
Brooke, J., D.  Bitko, T. F.  Rosenbaum and G. Aeppli,  Quantum
annealing of a disordered magnet. {\it Science} {\bf 284},
779--781 (1999).


\bibitem{Buyers86}
Buyers, W. J. L., R. Morra, R. L. Armstrong, M. J. Hogan, P.
Gerlach and K. Hirakawa, Experimental evidence for the Haldane gap
in a spin-1 nearly isotropic, antiferromagnetic chain. {\it Phys.
Rev. Lett.} {\bf 56} 371--374 (1986).


\bibitem{Darriet93}
Darriet, J. and L.-P. Regnault, The compound Y$_2$BaNiO$_5$: a new
example of a Haldane gap in a S=1 magnetic chain. {\it Solid State
Commun.}, {\bf 86}, 409--412 (1993).


\bibitem{Enderle99}
Enderle, M., Z.  Tun, W. J. L. Buyers and M. Steiner, Longitudinal
spin fluctuations of coupled integer-spin chains: Haldane triplet
dynamics in the ordered phase of CsNiCl$_3$. {\it Phys. Rev. B}
{\bf 59}, 4235--4243 (1999).


\bibitem{Garcia93}
Garc\'{i}a-Matres, E., J. L. Mart\'{i}nez, J.
Rodr\'{i}guez-Carjaval, J. A. Alonso, A. Salinas-S\'{a}nchez and
R. S\'{a}ez-Puche, Structural characterization and polymorphysm of
R$_2$BaNiO$_5$ (R=Nd, Gd, Dy, Y, Ho, Er, Tm, Yb) studied by
neutron diffraction, {\it J. Solid State Chem.} {\bf 103}, 322-333
(1993).

\bibitem{Garcia95}
Garc\'{i}a-Matres, E.,  J. L. Garc\'{i}a-Munoz, J. L.
Mart\'{i}nez, J. Rodriguez-Carvajal, Magnetic susceptibility and
field-induced transitions in R$_2$BaNiO$_5$ compounds (R=Tm, Er,
Ho, Dy, Tb, Gd, Sm, Nd, Pr).{\it J. Mag. Magn. Mater.} {\bf 149},
363--372 (1995).



\bibitem{Haldane83all}
Haldane, F. D. M., Continuum dynamics of the 1-D Heisenberg
antiferromagnet: identification with the O(3) nonlinear sigma
model. {\it Phys. Lett.} {\bf 93A}, 464--468 (1983); Non-linear
spin theory of large-spin Heisenberg antiferromagnets:
semiclassically quantized solitons of the one-dimensional
easy-axis Neel state. {\it Phys. Rev. Lett.} {\bf 50},
  1153--1156 (1983).

\bibitem{Jensen}
Jensen, J.  and Mackintosh, A. R., Rare Earth Magnetism, Clarendon
Press, Oxford, 1991.


\bibitem{Kim98}
Kim, Y. J., M. Greven, U.-J.  Wiese and  R, J, Birgeneau, Monte
carlo study of correlations in quantum spin chains at non-zero
temperature. {\it Eur. Phys. J. B} {\bf 4},  291  (1998).



\bibitem{Lake2000}
Lake, B., D. A. Tennant and S. E. Nagler, cond-mat/9910459.

\bibitem{numerics}
Lou, J.,  Dai, X.,  Quin, S.,  Su, Z, and Yu, L., Heisenberg
spin-one chain in staggered magnetic field : a density matrix
renormalization group study, {\it Phys Rev. B} {\bf 60}, 52--55
(1999).


\bibitem{MZ98-L}
Maslov, S. and Zheludev, A., Universal behavior of one-dimensional
antiferromagnets in staggered magnetic fields.  {\it Phys. Rev.
Lett.} {\bf 80}, 5786-5789 (1998).


\bibitem{Morra88}
Morra, R. M., W. J. L. Buyers, R. L. Armstrong and K. Hirakawa,
Spin dynamics and the Haldane gap in the spin-1
quasi-one-dimensional antiferromagnet CsNiCl$_3$. {\it Phys. Rev.
B} {\bf 38}, 543--555 (1988).

\bibitem{Raymond99}
Raymond, S., T. Yokoo, A. Zheludev, S. E. Nagler, A. Wildes and J.
Akimitsu, Polarized-neutron study of longitudinal Haldane-gap
excitations in Nd$_2$BaNiO$_5$. {\it Phys. Rev. Lett.} {\bf 82},
2382--2385 (1999).

\bibitem{Scalapino75}
Scalapino, D.~J., Y. Imry and  P. Pincus, Generalized
Ginzburg-Landau theory of pseudo-one-dimensional systems. {\it
Phys. Rev. B} {\bf 11}, 2042--2048 (1975).


\bibitem{Tennant93}
Tennant, D. A.,  T. G. Perring, R. A. Cowley and S. E. Nagler,
Unbound spinons in the S=1/2 antiferromagnetic chain KCuF$_3$.
{\it Phys. Rev. Lett.} {\bf 70}, 4003--4066 (1993).

\bibitem{Tennant95}
Tennant, D. A.,  S. E. Nagler, D. Welz, G. Shirane and  K. Yamada,
Effects of coupling between chains on the magnetic excitation
spectrum of KCuF$_3$, {\it Phys. Rev. B} {\bf 52}, 13381 (1995).

\bibitem{Xu96}
Xu, G., J. F. DiTusa, T. Ito, H. Takagi, C. L. Broholm and G.
Aeppli, Y$_2$BaNiO$_5$: a nearly ideal realization of the S=1
Heisenberg chain with antiferromagnetic interactions. {\it Phys.
Rev. B} {\bf 54}, R6827--R6830 (1996).



\bibitem{Yokoo95}
Yokoo, T., T. Sakaguchi, K. Kakurai,  and J. Akimitsu,  Neutron
scattering study of magnetic excitations in the spin S=1
one-dimensional antiferromagnet Y$_2$BaNiO$_5$. J. {\it J. Phys.
Soc. Japan}, {\bf 65}, 3025--3031, (1995).

\bibitem{Yokoo97}
Yokoo, T., A. Zheludev, M.  Nakamura and J. Akimitsu,  Ni-chain
gap excitations in (Nd$_x$Y$_{1-x}$)$_2$BaNiO$_5$: one-dimensional
to three-dimensional crossover.  {\it Phys. Rev. B} {\bf 55},
  11516--11520  (1997).


\bibitem{Yokoo98}
Yokoo, T., S. Raymond, A. Zheludev, S. Maslov, E. Ressouche, I.
Zaliznyak, R. Erwin, M. Nakamura and J. Akimitsu, Magnetic
ordering, spin waves, and Haldane-gap excitations in
(Nd$_x$Y$_{1-x}$)$_2$BaNiO$_5$ linear-chain mixed-spin
antiferromagnets.  {\it Phys. Rev. B} {\bf 58}, 14424--14435
(1998).


\bibitem{Zheludev96e}
Zheludev, A., J. M. Tranquada, T.  Vogt and D. J.  Buttrey,
Magnetic excitations and soft-mode transition in Pr$_2$BaNiO$_5$.
{\it Europhys. Lett.} {\bf 35},  385--390 (1996); Magnetic
excitations and soft-mode transition in the quasi-one-dimensional
mixed-spin antiferromagnet Pr$_2$BaNiO$_5$. {\it Phys. Rev. B}
{\bf 54}, 6437--6447 (1996).

\bibitem{Zheludev96}
Zheludev, A., J. M. Tranquada, T.  Vogt and D. J. Buttrey,
Magnetic gap excitations in a one-dimensional mixed-spin
antiferromagnet Nd$_2$BaNiO$_5$. {\it Phys. Rev. B} {\bf   54},
7210--7215 (1996).



\bibitem{Zheludev00}
Zheludev, A., S. Maslov, T. Yokoo, J. Akimitsu, S. Raymond, S. E.
Nagler and K. Hirota, Role of single-ion excitations in the
mixed-spin quasi-one-dimensional quantum antiferromagnet
Nd$_2$BaNiO$_5$, Phys. Rev. B {\bf 61}, 11601 (2000).

\bibitem{ZM98NBANO-L}
Zheludev, A.,  E. Ressouche, S. Maslov, T. Yokoo, S. Raymond  and
J. Akimitsu, Experimental measurement of the staggered
magnetization curve for a Haldane spin chain, {\it Phys. Rev.
Lett.} {\bf 80}, 3630--3633 (1998).





\end{thebibliography}
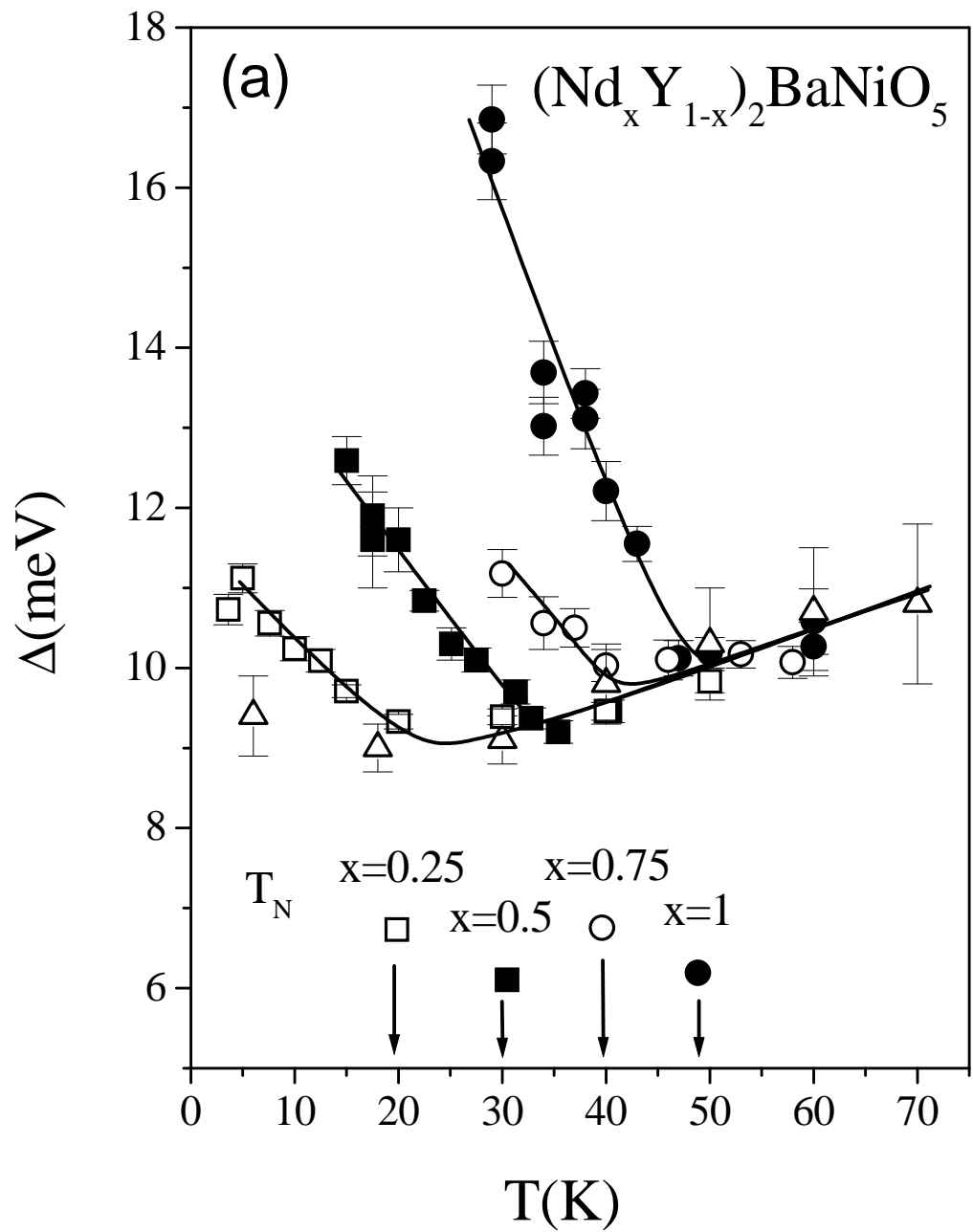
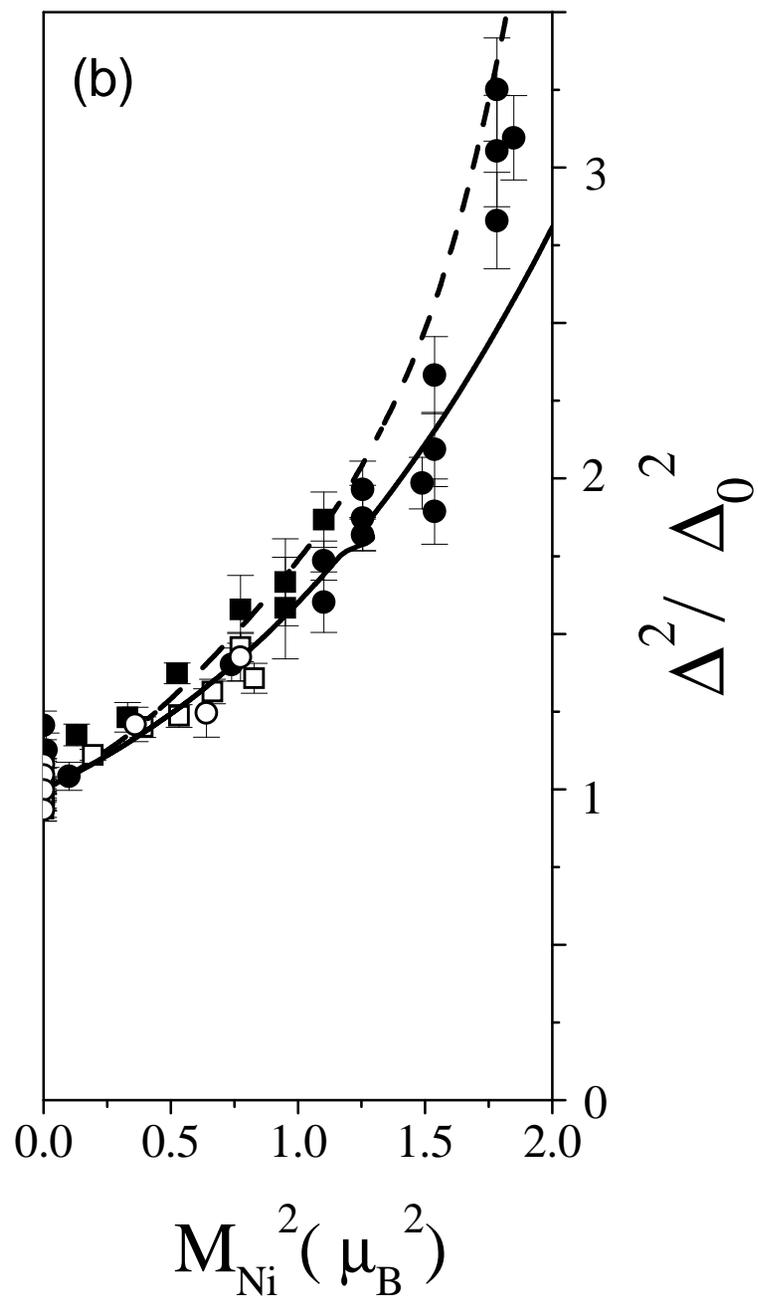
\end{document}